# General pulsed-field gradient signal attenuation expression based on a fractional integral modified-Bloch equation

Guoxing Lin*

*Carlson School of Chemistry and Biochemistry, Clark University, Worcester, MA 01610, USA*

**Abstract**

Anomalous diffusion has been investigated in many polymer and biological systems. The analysis of PFG anomalous diffusion relies on the ability to obtain the signal attenuation expression. However, the general analytical PFG signal attenuation expression based on the fractional derivative has not been previously reported. Additionally, the reported modified-Bloch equations for PFG anomalous diffusion in the literature yielded different results due to their different forms. Here, a new integral type modified-Bloch equation based on the fractional derivative for PFG anomalous diffusion is proposed, which is significantly different from the conventional differential type modified-Bloch equation. The merit of the integral type modified-Bloch equation is that the original properties of the contributions from linear or nonlinear processes remain unchanged at the instant of the combination. From the modified-Bloch equation, the general solutions are derived, which includes the finite gradient pulse width (FGPW) effect. The numerical evaluation of these PFG signal attenuation expressions can be obtained either by the Adomian decomposition, or a direct integration method that is fast and practicable. The theoretical results agree with the continuous-time random walk (CTRW) simulations performed in this paper. Additionally, the relaxation effect in PFG anomalous diffusion is found to be different from that in PFG normal diffusion. The new modified-Bloch equations and their solutions provide a fundamental tool to analyze PFG anomalous diffusion in nuclear magnetic resonance (NMR) and magnetic resonance imaging (MRI).

## 1. Introduction

Anomalous diffusion widely exists in polymer and biological system [1,2,3]. However, the study of anomalous diffusion by pulsed field gradient (PFG) [4,5,6] diffusion technique still faces challenges due to the complexity of PFG anomalous diffusion theory. Although there are many theoretical efforts to investigate the PFG anomalous diffusion [7,8,9,10,11,12,13,14,15,16,17,18,19,20], the general analytical PFG signal attenuation expression for anomalous diffusion based on the fractional derivative [21,22,23] has not been reported. Additionally, the modified-Bloch equations [14,16] for anomalous diffusion reported in the literature have different forms, and yield inconsistent theoretical results. Moreover, the short gradient pulse (SGP) approximation's result obtained in Ref. [18] is hard to obtain from these modified-Bloch equations. To better analyze PFG fractional diffusion, it may still be valuable to develop new modified-Bloch equations for PFG anomalous diffusion.

In studying PFG diffusion, the modified Bloch equation [24] proposed by Torrey in 1956 has been one of the most successful PFG theories for normal diffusion, which is a combination of different processes such as diffusion, Larmor precession, and relaxation. However, the combination of different processes becomes a challenge to build a modified-Bloch equation for PFG anomalous diffusion. Precession and relaxation processes have a time derivative $\frac{\partial}{\partial t}$, while fractional diffusion has a significantly different time derivative with the derivative order $0 < \alpha \leq 2$. To overcome the difficulty due to the different time derivatives, a new type modified-Bloch equation, an integral equation, is proposed. This integral type equation combines the contributions from different processes such as diffusion, precession and relaxation into an integral with the same time increment $dt$. Through the integral modified-Bloch equation, a general PFG signal attenuation equation is obtained, which can be used to evaluate PFG signal attenuations by a direct integration method numerically. On the other hand, the general PFG signal attenuation equation can be solved by the Adomian decomposition method to give an analytical PFG signal attenuation expression, which can be used to evaluate the PFG signal attenuation numerically as well. The direct integration method and the Adomian decomposition method get the same numerical results of PFG signal attenuation. The PFG signal attenuation obtained in this paper includes the finite gradient pulse width (FGPW) effect. Additionally, continuous-time random walk (CTRW) simulation [25] is carried out to verify these theoretical results. The CTRW simulation method employed here has been developed in Ref. [26], which is based on two models: the CTRW model proposed in Ref. [25] and the lattice model proposed in Refs. [27,28] Furthermore, the relaxation effect in PFG anomalous diffusion experiments is considered.

## 2. Theory

For the sake of simplicity, only one-dimensional free anomalous diffusion in a homogeneous sample is investigated here. Pulsed-field gradient technique [29,30] is an established non-invasive tool to measure diffusion in nuclear magnetic resonance (NMR) and magnetic resonance imaging (MRI). It applies gradient pulse to artificially create a time and space dependent magnetic field: $\mathbf{B}(\mathbf{z},t) = \mathbf{B}_0 + \mathbf{g}(t) \cdot \mathbf{z}$, where $\mathbf{B}_0$ is the exterior magnetic field, $\mathbf{g}(t)$ is the gradient vector, and $\mathbf{z}$ is the position vector. The spin moment precesses around the magnetic field at Larmor frequency $\boldsymbol{\omega}(\mathbf{z},t) = -\gamma\mathbf{B}(\mathbf{z},t)$, where $\gamma$ is the gyromagnetic ratio. In a rotating frame rotating around a magnetic field at angular frequency $\boldsymbol{\omega}_0 = -\gamma\mathbf{B}_0$, the accumulating phase $\phi(t)$ (the net phase change of spin precession induced by the gradient pulses) can be written as [29,30]

$$\phi(t) = -\int_0^t \gamma|\boldsymbol{\omega}(\mathbf{z},t') - \boldsymbol{\omega}_0|dt' = -\int_0^t \gamma\mathbf{g}(t') \cdot \mathbf{z}(t')dt'. \tag{1}$$

Eq. (1) is a path integral. In a diffusion spin system, the spins undergo numerous possible paths and the accumulated phase will belong to either a Gaussian or a non-Gaussian distribution. The NMR signal comes from the average signal from spins with all possible phases, which can be described as

$$S(t) = S(0)\int_{-\infty}^{\infty} P(\phi,t)\exp(+i\phi)d\phi, \tag{2}$$

where $S(0)$ is the signal intensity at the beginning of the first dephasing gradient pulse, $S(t)$ is the signal intensity at time $t$, and $P(\phi,t)$ is the phase probability distribution function (only the symmetric $P(\phi,t)$ is focused in this paper). In this paper, normalized signal intensity, namely $S(0) = 1$, will be used, and $S(t)$ is the PFG signal attenuation.

Different theoretical methods analyze PFG diffusion in different ways. Some methods are based on Eqs. (1) and (2), which includes Gaussian or non-Gaussian phase approximation method [7,19], or the effective phase shift diffusion equation method [18], etc. In these kinds of methods, the phase distribution $P(\phi,t)$ results from each individual spin's accumulating phase described by Eq. (1). While some other methods describe the PFG diffusion in a significantly different way. They do not require the knowledge of $P(\phi,t)$ or the phase path integral, but rather focus on the evolution of local spin magnetization $M_{xy}(\mathbf{z},t) = M_x(\mathbf{z},t) + iM_y(\mathbf{z},t)$. The local magnetization $M_{xy}(\mathbf{z},t)$ is as an average magnetization contributed from all spins in position $z$ at the instant of time $t$, which is an ensemble effect of local spins. Similar to the Larmor precession of spin moment, the magnetization $M_{xy}(\mathbf{z},t)$ precesses around the magnetic field $\mathbf{B}(\mathbf{z},t) = \mathbf{B}_0 + \mathbf{g}(t) \cdot \mathbf{z}$ by a time and space dependent angular speed, $\boldsymbol{\omega}(\mathbf{z},t) = -\gamma\mathbf{B}(\mathbf{z},t)$. Therefore, the magnetization $M_{xy}(\mathbf{z},t)$ has a time and space dependent phase. At each instant of the diffusion, magnetizations with different phases diffusing to the same spot will mix together, resulting in an instantaneous signal attenuation at the mixing spot. The total PFG signal attenuation is the accumulating effect of magnetization diffusion. Various methods can describe the signal attenuation due to the mixing of magnetization in diffusion. These methods include the modified-Bloch equation [24], the instantaneous signal attenuation method [26], etc. The instantaneous signal attenuation method does not need the path integral. The modified-Bloch equation method [24] does not rely on both the $P(\phi,t)$ and the phase path integral. These two methods have advantages in systems where the phase evolution information could be hard to obtain such as in inhomogeneous system, restricted diffusion, and nonlinear gradient field. The modified-Bloch equation is one of the most fundamental PFG theoretical methods, and it will be the focus of this paper. The subsequent section will show how to build the modified-Bloch equation for PFG anomalous diffusion.

*2.1 The modified-Bloch equation*

In PFG diffusion experiments [29,30], the spin magnetization evolution is simultaneously affected by different processes such as Larmor precession, relaxation, and diffusion. The spin magnetization on XY plane [29,30], $M_{xy}(\mathbf{z},t) = M_x(\mathbf{z},t) + iM_y(\mathbf{z},t)$ presseses under the influence of gradient magnetic field, $\mathbf{B}(\mathbf{z},t) = \mathbf{B}_0 + \mathbf{g}(t) \cdot \mathbf{z}$. In a rotating frame rotating around a magnetic field at angular frequency $\boldsymbol{\omega} = -\gamma \mathbf{B}_0$, the spin system precession can be described by [29,30]

$$\frac{\partial}{\partial t} M_{xy}(\mathbf{z},t) = -i\gamma \mathbf{g}(t) \cdot \mathbf{z} M_{xy}(\mathbf{z},t). \tag{3}$$

The relaxation of the transverse component of magnetization $M_{xy}(\mathbf{z},t)$ can be described as [31,32]

$$\frac{\partial}{\partial t} M_{xy}(\mathbf{z},t) = -\frac{M_{xy}(\mathbf{z},t)}{T_2}, \tag{4}$$

where $T_2$ is the spin-spin relaxation time constant. The modified-Bloch equation for normal diffusion is [24,33]

$$\frac{\partial}{\partial t} M_{xy}(\mathbf{z},t) = D \frac{\partial^2}{\partial z^2} M_{xy}(\mathbf{z},t) - i\gamma \mathbf{g}(t) \cdot \mathbf{z} M_{xy}(\mathbf{z},t) - \frac{M_{xy}(\mathbf{z},t)}{T_2}, \tag{5}$$

a straightforward combination of the normal diffusion, Larmor precession and relaxation equation. Such a straightforward combination works because these processes have the same time derivative $\frac{\partial}{\partial t}$. However, compared to $\frac{\partial}{\partial t}$, the time fractional derivative of fractional diffusion has a different derivative order $\alpha$, $0 < \alpha \le 2$, and a remarkably different form. It may not be appropriate to directly combine the fractional diffusion equation with the precession and relaxation equations.

The time-space fractional diffusion could be modeled by the fractional derivative as [21-23]

$$_tD_*^\alpha M_{xy}(\mathbf{z},t) = D_f \frac{\partial^\beta}{\partial |z|^\beta} M_{xy}(\mathbf{z},t), \tag{6}$$

where $D_f$ is the fractional diffusion coefficient with units $m^\beta / s^\alpha$, $\frac{\partial^\beta}{\partial |z|^\beta}$ is the space fractional derivative defined in Appendix A, and $_tD_*^\alpha$ is the Caputo fractional derivative defined as [21-23]

$$_tD_*^\alpha f(t) := \begin{cases} \dfrac{1}{\Gamma(m-\alpha)} \int_0^t \dfrac{f^{(m)}(\tau) d\tau}{(t-\tau)^{\alpha+1-m}}, & m-1 < \alpha < m, \\ \dfrac{d^m}{dt^m} f(t), & \alpha = m. \end{cases}$$

The Caputo time derivative $_tD_*^\alpha$ is apparently different from $\frac{\partial}{\partial t}$, which makes it difficult to combine the fractional diffusion equation with the precession equation.

Nevertheless, the combination can be carried out in a different way. The Caputo fractional derivative has the following

property [34]:

$$J^\alpha \,_t D_*^\alpha \mu(t) = u(t) - \sum_{k=0}^{m-1} u^{(k)}(0^+) \frac{t^k}{k!}, \tag{7}$$

where $J^\alpha(f(t)) = \int_0^t \frac{(t-\tau)^{\alpha-1}}{\Gamma(\alpha)} f(\tau) d\tau$ [21,22]. By operating $J^\alpha$ on both sides of the fractional diffusion equation, Eq. (6), we get

$$M_{xy}(\mathbf{z},t) = \sum_{k=0}^{m-1} M^{(k)}(z,0^+) \frac{t^k}{k!} + J^\alpha \left[ D_f \frac{\partial^\beta}{\partial |z|^\beta} M_{xy}(\mathbf{z},t) \right], \tag{8a}$$

which is equivalent to

$$M(z,t) = \sum_{k=0}^{m-1} M^{(k)}(z,0^+) \frac{t^k}{k!} + \int_0^t \frac{(t-\tau)^{\alpha-1}}{\Gamma(\alpha)} D_f \frac{\partial^\beta}{\partial |z|^\beta} M_{xy}(\mathbf{z},t) d\tau. \tag{8b}$$

By operating $\frac{\partial}{\partial t}$ on both side of Eqs. (8a) and (8b), we get

$$\begin{aligned}
\frac{d}{dt} M(z,t) &= \frac{d}{dt} \left[ \sum_{k=0}^{m-1} M^{(k)}(z,0^+) \frac{t^k}{k!} + \int_0^t \frac{(t-\tau)^{\alpha-1}}{\Gamma(\alpha)} D_f \frac{\partial^\beta}{\partial |z|^\beta} M_{xy}(\mathbf{z},\tau) d\tau \right] \\
&= \frac{d}{dt} \left[ \int_0^t \frac{(t-\tau)^{\alpha-1}}{\Gamma(\alpha)} D_f \frac{\partial^\beta}{\partial |z|^\beta} M_{xy}(\mathbf{z},\tau) d\tau \right]
\end{aligned} \tag{9}$$

where $\frac{\partial}{\partial t} M_{xy}(\mathbf{z},0^+) = 0$ is utilized. This condition $\frac{\partial}{\partial t} M_{xy}(\mathbf{z},0^+) = 0$ has been suggested in the Refs. [21] and [23] to get a continuous transition from $\alpha = 1^-$ to $\alpha = 1^+$, and it will be employed in the rest of this paper. At $\beta = 2$, Eq. (9) belongs to the same type anomalous diffusion equation obtained from a heuristic derivation in Ref. [35]. The precession equation, Eq. (3) can be rewritten as

$$\begin{aligned}
\frac{d}{dt} M_{xy}(\mathbf{z},t) &= \frac{d}{dt} \left[ M_{xy}(\mathbf{z},0) - \int_0^t i\gamma \mathbf{g}(\tau) \cdot \mathbf{z} M_{xy}(\mathbf{z},\tau) d\tau \right] \\
&= \frac{d}{dt} \left[ -\int_0^t i\gamma \mathbf{g}(\tau) \cdot \mathbf{z} M_{xy}(\mathbf{z},\tau) d\tau \right],
\end{aligned} \tag{10}$$

where $M_{xy}(\mathbf{z},0)$ is a constant. In Eq. (10), the gradient field rotates the local magnetization by an angle $\gamma \mathbf{g}(\tau) \cdot \mathbf{z} d\tau$ during $d\tau$. By combining Eqs. (9) and (10), we have

$$\frac{d}{dt} M_{xy}(\mathbf{z},t) = \frac{d}{dt} \int_0^t \left[ \frac{(t-\tau)^{\alpha-1}}{\Gamma(\alpha)} D_f \frac{\partial^\beta}{\partial |z|^\beta} M_{xy}(\mathbf{z},\tau) - i\gamma \mathbf{g}(\tau) \cdot \mathbf{z} \cdot M_{xy,\tau}(\mathbf{z},t) \right] d\tau,$$

or an equivalent equation,

$$M_{xy}(\mathbf{z},t) = \sum_{k=0}^{m-1} M_{xy}^{(k)}(\mathbf{z},0^+) \frac{t^k}{k!} + \int_0^t \left[ \frac{(t-\tau)^{\alpha-1}}{\Gamma(\alpha)} D_f \frac{\partial^\beta}{\partial |z|^\beta} M_{xy}(\mathbf{z},\tau) - i\gamma \mathbf{g}(\tau) \cdot \mathbf{z} \cdot M_{xy,\tau}(\mathbf{z},t) \right] d\tau, \tag{11}$$

where

$$M_{xy,\tau}(\mathbf{z},t) = \sum_{k=0}^{m-1} M_{xy}{}^{(k)}(\mathbf{z},0^+)\frac{t^k}{k!} + \int_0^\tau \left[\frac{(t-\tau')^{\alpha-1}}{\Gamma(\alpha)} D_f \frac{\partial^\beta}{\partial |z|^\beta} M_{xy}(\mathbf{z},\tau') - i\gamma \mathbf{g}(\tau')\cdot \mathbf{z}\cdot M_{xy,\tau'}(\mathbf{z},t)\right] d\tau',$$

is the partially calculated $M_{xy}(\mathbf{z},t)$ value at time $\tau$ staying in position $\mathbf{z}$ that will be detected in the final time $t$. When $\alpha = 1$, $M_{xy,\tau}(\mathbf{z},t) = M_{xy}(\mathbf{z},\tau)$, while when $\alpha \neq 1$, $M_{xy,\tau}(\mathbf{z},t)$ is different from $M_{xy}(\mathbf{z},\tau)$. Eq. (11) is the modified-Bloch equation built upon the fractional derivative. The gradient field rotates both the magnetizations $M_{xy,\tau}(\mathbf{z},t)$ and $M_{xy}(\mathbf{z},\tau)$ by an angle $\gamma \mathbf{g}(\tau)\cdot \mathbf{z}\, d\tau$ during time $d\tau$. However,

$$M_{xy,\tau}(\mathbf{z},t) - i\gamma \mathbf{g}(\tau)\cdot \mathbf{z} M_{xy,\tau}(\mathbf{z},t) d\tau = \exp[-i\gamma \mathbf{g}(\tau)\cdot \mathbf{z} d\tau] M_{xy,\tau}(\mathbf{z},t),$$

while because $M_{xy,\tau}(\mathbf{z},t) \neq M_{xy}(\mathbf{z},\tau)$,

$$M_{xy,\tau}(\mathbf{z},t) - i\gamma \mathbf{g}(\tau)\cdot \mathbf{z} M_{xy}(\mathbf{z},\tau) d\tau \neq \exp[-i\gamma \mathbf{g}(\tau)\cdot \mathbf{z} d\tau] M_{xy,\tau}(\mathbf{z},t), \text{when } \alpha \neq 1.$$

Therefore, $-i\gamma \mathbf{g}(\tau)\cdot \mathbf{z} M_{xy}(\mathbf{z},\tau)d\tau$ will not give the proper phase rotation in Eq. (11) and only the rotation of $M_{xy,\tau}(\mathbf{z},t)$ that affects the calculation of magnetization $M_{xy}(\mathbf{z},t)$ will be considered in Eq. (11). In Eq. (11), the term $\frac{(t-\tau)^{\alpha-1}}{\Gamma(\alpha)} D_f \frac{\partial^\beta}{\partial |z|^\beta} M_{xy}(\mathbf{z},t) d\tau$ is the diffusion related attenuation during the interval $d\tau$ that will affect the $M_{xy}(\mathbf{z},t)$ at time $t$. During the same interval, the precession alters the magnetization $M_{xy,\tau}(\mathbf{z},t)$ by $-i\gamma \mathbf{g}(\tau)\cdot \mathbf{z}\cdot M_{xy,\tau}(\mathbf{z},t) d\tau$. Eq. (11) is significantly different from the modified-Bloch equation proposed in references [14,16].

For a homogeneous sample, the magnetization is $M_{xy}(\mathbf{z},\tau) = S(\tau)\exp(-i\mathbf{K}(\tau)\cdot \mathbf{z})$, where $\mathbf{K}(\tau) = \int_0^\tau \gamma \mathbf{g}(t')dt'$ is the wavenumber [29,30]. By substituting $M_{xy}(\mathbf{z},\tau) = S(\tau)\exp(-i\mathbf{K}(\tau)\cdot \mathbf{z})$ into Eq. (11), we can get

$$S(t)\exp(-i\mathbf{K}(t)\cdot \mathbf{z}) = \sum_{k=0}^{m-1} M_{xy}{}^{(k)}(\mathbf{z},0^+)\frac{t^k}{k!} + \int_0^t \left[-\frac{(t-\tau)^{\alpha-1}}{\Gamma(\alpha)} D_f |K(\tau)|^\beta S(\tau)\exp(-i\mathbf{K}(\tau)\cdot \mathbf{z}) - i\gamma \mathbf{g}(\tau)\cdot \mathbf{z}\cdot M_{xy,\tau}(\mathbf{z},t)\right] d\tau,$$

which reduces to the following two equivalent equations at the origin where $\mathbf{z}=0$ and $\exp(-i\mathbf{K}(t)\cdot \mathbf{z})=1$ in a homogeneous system:

$${}_tD_*^\alpha S(t) = -D_f |\mathbf{K}(\tau)|^\beta S(t), \tag{12a}$$

$$S(t) = \sum_{k=0}^{m-1} S^{(k)}(0^+)\frac{t^k}{k!} - J^\alpha\left(D_f |\mathbf{K}(\tau)|^\beta S(\tau)\right). \tag{12b}$$

Other three different ways to derive the Eq. (12) from Eq. (11) can be seen in Appendix B and Appendix C, where very detailed derivations are given.

It is difficult to add the relaxation effect to the PFG fractional diffusion based on the fractional derivative, which may be approached in two ways:

i. the relaxation is not relevant to the anomalous diffusion and precession, then $M_{xy}(\mathbf{z},\tau) = S(\tau)\exp(-\tau/T_2)\exp(-i\mathbf{K}(\tau)\cdot \mathbf{z})$, and from Eq. (11) we have

$$M_{xy}(\mathbf{z},t) = \exp\left(-\frac{t}{T_2}\right)\left\{\sum_{k=0}^{m-1} M_{xy}{}^{(k)}(\mathbf{z},0^+)\frac{t^k}{k!} + \int_0^t \left[\frac{(t-\tau)^{\alpha-1}}{\Gamma(\alpha)} D_f \frac{\partial^\beta}{\partial |z|^\beta} M_{xy}(\mathbf{z},\tau) - i\gamma \mathbf{g}(\tau)\cdot \mathbf{z}\cdot M_{xy,\tau}(\mathbf{z},t)\right] d\tau\right\}.$$

(13)

Eq. (13) will give a signal attenuation that is proportional to $\exp(-t/T_2)$.

ii. the relaxation depends on the instantaneous signal intensity affected by the PFG fractional diffusion. In this case, the relaxation Eq. (4) can be written as $\frac{d}{dt}M(z,t) = \frac{d}{dt}\left\{M_{xy}(\mathbf{z},0) - \int_0^t \left(\frac{1}{T_2}M_{xy}(\mathbf{z},\tau)\right)d\tau\right\}$ which can be added to the integral in Eq. (11) to give

$$M_{xy}(\mathbf{z},t) = \sum_{k=0}^{m-1} M_{xy}^{(k)}(\mathbf{z},0^+)\frac{t^k}{k!} + \int_0^t \left[\frac{(t-\tau)^{\alpha-1}}{\Gamma(\alpha)}D_f \frac{\partial^\beta}{\partial |z|^\beta}M_{xy}(\mathbf{z},\tau) - \frac{1}{T_2}M_{xy}(\mathbf{z},\tau) - i\gamma \mathbf{g}(\tau)\cdot\mathbf{z}\cdot M_{xy,\tau}(\mathbf{z},t)\right]d\tau,$$

(14)

the integral type modified-Bloch equation including the $T_2$ relaxation. For the free diffusion in a homogeneous system, Eq. (14) reduces to

$$S(t) = \sum_{k=0}^{m-1} S^{(k)}(0)\frac{t^k}{k!} + \int_0^t \left\{\left(\frac{(t-\tau)^{\alpha-1}}{\Gamma(\alpha)}D_f K^\beta(t) - \frac{1}{T_2}\right)S(\tau)\right\}d\tau.$$

(15)

When $\alpha \neq 1$, Eq. (15) will give a signal attenuation that is not proportional to $\exp(-t/T_2)$.

When $\alpha = 1, \beta = 2$, the integral equation, Eq. (14) is equivalent to the traditional differential modified-Bloch equation for normal diffusion Eq. (5). For space fractional diffusion, whose $\alpha = 1$ and $0 < \beta \leq 2$, both Eqs. (13) and (14) can be rewritten as

$$\frac{\partial}{\partial t}M_{xy}(\mathbf{z},t) = \left[D_f \frac{\partial^\beta}{\partial |z|^\beta} - i\gamma \mathbf{g}(t)\cdot\mathbf{z} - \frac{1}{T_2}\right]M_{xy}(\mathbf{z},t),$$

(16)

which is consistent with the results reported in Refs. [14,16]. In a homogeneous system, Eq. (16) reduces to

$$\frac{\partial}{\partial t}S(t) = \left[-K^\beta(t)D_f - \frac{1}{T_2}\right]S(t),$$

(17)

which yields the PFG signal attenuation

$$S(t) = \exp(-\frac{1}{T_2})\exp\left[-D_f(\gamma g\delta)^\beta(\Delta - \frac{\beta-1}{\beta+1}\delta)\right].$$

(18)

Eq. (18) agrees with the results reported from both modified-Bloch equations in the Refs. [14] and [16].

*2.2 Analytical PFG signal attenuation expression obtained by the Adomian decomposition method*

Eqs. (12a) and (12b) describe the PFG signal attenuation in free general fractional diffusion with $\{0 < \alpha, \beta \leq 2\}$, including time-fractional diffusion, space-fractional diffusion and normal diffusion. The similar type of fractional equation as Eqs. (12b) has been solved by the Adomian decomposition Method [34,36,37,38]. According to the results from these references, the solution of Eq. (12b) is [34,38]

$$S(t) = \sum_{n=0}^{\infty} S_n(t),$$

(19a)

where

$$S_0(t) = \sum_{k=0}^{m-1} S^{(k)}(0^+)\frac{t^k}{k!}, m-1 < \alpha < m,$$

(19b)

and

$$S_n(t) = \int_0^t \left\{ \left( -\frac{(t-\tau)^{\alpha-1}}{\Gamma(\alpha)} D_f K^\beta(t) \right) S_{n-1}(\tau) \right\} d\tau . \quad (19c)$$

When the relaxation effect is considered but it is independent to the fractional diffusion, based on Eqs. (13) and (19c), we have

$$S_n(t) = \exp\left(-\frac{t}{T_2}\right) \int_0^t \left\{ \left( -\frac{(t-\tau)^{\alpha-1}}{\Gamma(\alpha)} D_f K^\beta(t) \right) S_{n-1}(\tau) \right\} d\tau . \quad (19d)$$

While, based on Eqs. (14), (17) and (19c), if the relaxation is relevant to the fractional diffusion, we have

$$S_n(t) = \int_0^t \left\{ \left( -\frac{(t-\tau)^{\alpha-1}}{\Gamma(\alpha)} D_f K^\beta(t) - \frac{1}{T_2} \right) S_{n-1}(\tau) \right\} d\tau . \quad (19e)$$

Future experimental research is required to test which one of the two equations, Eqs. (19d) and (19e), is consistent with experimental results.

### 2.3. The direct integration method for numerical evaluation of Mittag-Leffler type function based PFG signal attenuation

Although the analytical expressions, Eqs. (19a-e) obtained by the Adomian decomposition method can be used to numerically evaluate the PFG signal attenuation, its speed is not fast enough, and it may have overflow problem in large attenuation data because it requires superposition of many terms. These shortcomings may limit its potential practical applications. For instance, PFG MRI experiments usually include large data, which really need a fast numerical method to analyze. In this section, a practicable numerical evaluation method, a direct integration method [39] is employed for the numerical evaluation. From Eqs. (15), when we set $\sum_{k=0}^{m-1} S^{(k)}(0) \frac{t^k}{k!} = 1$, we have

$$\begin{aligned} S(t) &= 1 - \int_0^t \left\{ \left( \frac{(t-\tau)^{\alpha-1}}{\Gamma(\alpha)} D_f K^\beta(t) + \frac{1}{T_2} \right) S(\tau) \right\} d\tau \\ &= 1 - \int_0^t \left[ D_f K^\beta(t) S(\tau) \frac{d(t-\tau)^\alpha}{\Gamma(1+\alpha)} + \frac{1}{T_2} S(\tau) d\tau \right] \end{aligned}. \quad (20)$$

Examine the Eq. (20) closely, the PFG signal intensity $S(t)$ depends on the historical PFG signal intensities, $S(\tau)$, $0 \le \tau < t$. Therefore, the signal intensity can be discretely calculated step-by-step beginning from $t = 0$. By dividing the time into small intervals, at time $t_j$, $t_j = \sum_{k=1}^{j} \Delta t_k$, Eq. (20) can be rewritten in a discrete form as

$$S(t_j) = 1 - \sum_{k=1}^{j} S(t_{k-1}) \left\{ D_f K^\beta(t) [(t_j - t_{k-1})^\alpha - (t_j - t_k)^\alpha] / \Gamma(1.+\alpha) + \frac{1}{T_2}(t_k - t_{k-1}) \right\} . \quad (21)$$

Neglecting $T_2$ relaxation, Eq. (21) reduces to

$$S(t_j) = 1 - \sum_{k=1}^{j} a(t_k) S(t_{k-1}) [(t_j - t_{k-1})^\alpha - (t_j - t_k)^\alpha] / \Gamma(1.+\alpha), \quad (22)$$

where $a(t_k) = D_f K^\beta(t_k)$. From Eqs. (21) and (22), the PFG signal attenuations: $S(t_1)$, $S(t_2)$,..., $S(t_n)$ can be calculated step by step starting from $S(t_1)$ to $S(t_n)$.

Furthermore, from Eq. (22), if we denote $E_{\alpha,1,a(t)}(-t) = s(t)$, we get

$$E_{\alpha,1,a(t)}(-t) = 1 - \int_0^t \left\{ \left( \frac{(t-\tau)^{\alpha-1}}{\Gamma(\alpha)} a(\tau) \right) E_{\alpha,1,a(t)}(-t) \right\} d\tau, \quad (23)$$

where

$$E_{\alpha,1,a(t)}(-t) = \begin{cases} E_{\alpha,1}(-t^\alpha), & a(t) = 1 \\ E_{\alpha,1}(-ct^\alpha), & a(t) = c, c > 0 \end{cases}, \quad (24)$$

where $c$ is a real constant. From Eqs. (24) and (25), the Mittag-Leffler type function $E_{\alpha,1}(-ct^\alpha)$ and its derivative can be numerically evaluated by:

$$E_{\alpha,1}(-ct_j^\alpha) = 1 - \sum_{k=1}^{j} cE_{\alpha,1}(-ct_{k-1}^\alpha)\left[(t_j - t_{k-1})^\alpha - (t_j - t_k)^\alpha\right]/\Gamma(1+\alpha), \quad (25)$$

$$E'_{\alpha,1}(-ct_j^\alpha) = \frac{\left[E_{\alpha,1}(-ct_j^\alpha) - E_{\alpha,1,a(t)}(-ct_{j-1}^\alpha)\right]}{\Delta t_j}. \quad (26)$$

The direct integration method gives the same PFG signal attenuation as that obtained by the Adomian decomposition method, but it is much faster and does not cause overflow in the calculation.

## 3. CTRW simulation in a lattice model

CTRW simulations were performed to verify the theoretical results obtained in this paper. The PFG signal attenuation CTRW simulation method employed in this paper has been developed in Ref. [26], which is based on two models, the CTRW model [25] and the Lattice model [27,28]. The CTRW consists of a sequence of independent random waiting time and jump length combinations $(\Delta t_1, \Delta \xi_1)$, $(\Delta t_2, \Delta \xi_2)$, $(\Delta t_3, \Delta \xi_3)$,..., $(\Delta t_n, \Delta \xi_n)$. The individual waiting time $\Delta t$ and jump length $\Delta \xi$, are produced according to Ref. [25] by

$$\Delta t = -\eta_t \log U \left(\frac{\sin(\alpha_{i'}\pi)}{\tan(\alpha_{i'}\pi V)} - \cos(\alpha \pi)\right)^{\frac{1}{\alpha_{i'}}}, \quad (27)$$

and

$$\Delta \xi = \eta_z \left(\frac{-\log U \cos(\Phi)}{\cos((1-\beta)\Phi)}\right)^{1-\frac{1}{\beta}} \frac{\sin(\beta\Phi)}{\cos(\Phi)}, \quad (28)$$

where $\eta_t$ and $\eta_z$ are scale constants, $\Phi = \pi(V - 1/2)$, and $U, V \in (0,1)$ are two independent random numbers. Ref. [25] pointed out that the CTRW simulation based on the above waiting time and jump length can satisfy the time-space fractional diffusion equation under the diffusive limit, allowing the CTRW model to simulate anomalous diffusion in various academic fields, such as physics and economics.

The Lattice model developed in Refs. [27,28] was modified to record the spin phase. This lattice model has been applied to simulate PFG diffusion in polymer system [40]. The spin phase in the simulation is

$$\phi_i(t) = \sum_{j=1}^{n} \gamma g(t_j) z(t_j) \Delta t_j, \quad (29)$$

where $\phi_i(t)$ is the phase of the $i^{\text{th}}$ walk, $t_j = \sum_{k=1}^{j} \Delta t_k$, and $z(t_j) = \sum_{k=1}^{j} \Delta \xi_k$. The PFG signal attenuation in the simulation is the average over all the walkers in the simulation [28]

$$S(t) = \langle \cos[\phi_i(t)] \rangle = \frac{1}{N_{walks}} \sum_{i=1}^{N_{walks}} \cos[\phi_i(t)], \tag{30}$$

where $N_{walks}$ is the total number of the walks. A total of 100,000 walks were carried out for each simulation.

As the CTRW model [25] is proposed only for subdiffusion simulation, the simulation results here are limited to the subdiffusion. Interested readers are referred to Refs. [25-28] for more detailed information.

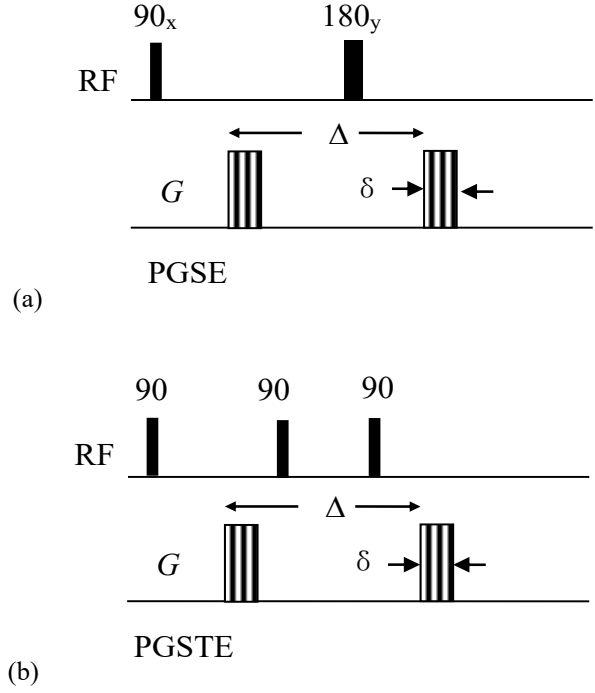

**Fig. 1** (a) PGSE pulse sequences, (b) PGSTE pulse sequence. The commonly used gradient pulse intensity *g* ranges from a few gauss/cm to hundreds of gauss/cm, and its width $\delta$ ranges from a few milliseconds to several tens milliseconds. The diffusion delay Δ is the time from the beginning of the dephasing gradient pulse to the beginning of the rephasing gradient pulse.

## 4. Discussion

In summary, an integral type of modified-Bloch equation was built for PFG fractional diffusion, and the general PFG signal attenuation expression for free fractional diffusion was derived. The obtained PFG signal attenuation expressions, Eqs. (19a)-(19d), are Mittag-Leffler type Function. The typical pulsed gradient spin echo (PGSE) and pulsed gradient stimulated echo (PGSTE) pulse sequences shown in Fig. 1 can be divided into three periods: $0 < t \leq \delta$, $\delta < t \leq \Delta$, and $\Delta < t \leq \Delta + \delta$ [29,30]. Neglecting the relaxation, for the free anomalous diffusion in a homogeneous sample, we obtain the following:

i. in Eq. (19b), if the $S^{(1)}(0^+) = 0$ condition is used [21,23], under SGP approximation we obtain

$$S(\Delta) = \sum_{n=0}^{\infty} S_n(t) = E_{\alpha,1}\left(-D_f K_{SGP}^{\beta} \Delta^{\alpha}\right), K_{SGP} = \gamma g \delta, \tag{31}$$

ii. for a single pulse attenuation, an ideal situation where the first gradient pulse is a regular pulse that commonly used in practical PFG experiments, which has a gradient intensity ranges from a few gauss/cm to a few hundred gauss/cm, and a time duration varies from a few miliseconds to tens millseconds. However, the second gradient pulse is infinitely narrow. In such a case, we get $S(t) = E_{\alpha,1+\beta/\alpha,\beta/\alpha}\left(-D_f(\gamma g)^\beta t^{\alpha+\beta}\right)$, where $E_{\alpha,\eta,\gamma}(x) = \sum_{n=0}^{\infty} c_n x^n, c_0 = 1, c_n = \prod_{k=0}^{n-1} \frac{\Gamma((k\eta+\gamma)\alpha+1)}{\Gamma((k\eta+\gamma+1)\alpha+1)}$ is a Mittag-Leffler type function, which is consistent with the results in Ref. [16]. The agreement is because Eq. (B.6), ${}_tD_*^\alpha\left[\exp(i\mathbf{K}(t)\cdot\mathbf{z})M_{xy}(\mathbf{z},t)\right] = \exp(i\mathbf{K}(t)\cdot\mathbf{z})D_f\frac{\partial^\beta}{\partial|z|^\beta}M_{xy}(\mathbf{z},t)$ can be derived from Eq. (11) for a homogeneous spin system, and Eq. (B.6) agrees with Ref. [16].

iii. at small PFG signal attenuation, Eqs. (19a)-(19c) agree with $S(t) = E_{\alpha,1}\left(-\int_0^t D_f K^\beta(t')dt'^\alpha\right)$ obtained by the instantaneous signal attenuation method [26] when only the first two terms are kept.

The CTRW simulations agree with the theoretical results as shown in Figs. 2 and 3. The relaxation effect is neglected in the theoretical results. Figs. (2a) and (2b) show how to obtain the fractional diffusion constant $D_f$. The fractional diffusion constant $D_f$ and other parameters such as $g$, $\delta$, $\alpha$, and $\beta$ are needed when using Eq. (19c) to calculate the theoretical PFG signal attenuation. In Fig. 2(b), $\langle z^\beta(t)\rangle$ with $\beta=1.5$ obtained from the simulation is finite, which is consistent with other reports [41,42]. In the reported CTRW simulation in Ref. [43], for a random walker in an imaginary growing box with a time-dependent spatial interval $L_1 t^{1/\beta} - L_2 t^{1/\beta}$, a finite $\langle z(t)^2\rangle_L = \int_{L_1 t^{1/\beta}}^{L_2 t^{1/\beta}} P(z,t)z^2 dz \sim t^{2\alpha/\beta}$ has been obtained. The $\langle z^\beta(t)\rangle$ with $\beta<2$ can be infinite in simulation, but the probabilities of a walker that has an infinite long jump in all the walkers during the limit time of the simulation are very small. Neglecting these small percentage of infinite long walkers will not obviously change the total signal intensity in most PFG experiments, where the signal is contributed by an enormous amount of spins. Additionally, even if $\langle z^\beta(t)\rangle$ with $\beta<2$ is infinite in theory, from expressions Eqs. (19a)-(19e), the range of normalized absolute value of PFG signal attenuation $|S(t)|$ is $0 \leq |S(t)| \leq 1$, which means that the signal intensity is finite and measurable in PFG experiments. In Figures 3(a) and 3(b), the derivative order parameters are $\alpha=0.6, \beta=2$, and $g$ equals 0.1 $T/m$. The $\alpha=0.6$ is chosen because it is near the $\alpha$ value of an ideal curvilinear diffusion. An ideal curvilinear diffusion has $\alpha=0.5, \beta=2$ [8,20,31,44]. For non-ideal curvilinear diffusion, it may have $\alpha \neq 0.5, \beta \neq 2$ [20]. In real applications, $\alpha=0.5, \beta=2$ has been used in Ref. [44] for the diffusion in micelles, while $\alpha=0.76, \beta=2$ has been reported in Ref. [31] for the curvilinear diffusion in a hydrated protein aerogel. In Fig. 3 (c), the derivative order parameters are $\alpha=0.75, \beta=1.5$, and $g$ equals 0.05 $T/m$ are used. Figs. 3(a) and 3(c) show that there is good agreement between the theoretical results and CTRW simulations at various $\Delta-\delta$ values. Fig. 3 (b) shows the agreement between the theoretical result and CTRW simulation under SGP approximation, where $\gamma g \delta = 6 \times 2.6751 \times 10^4$ is used.

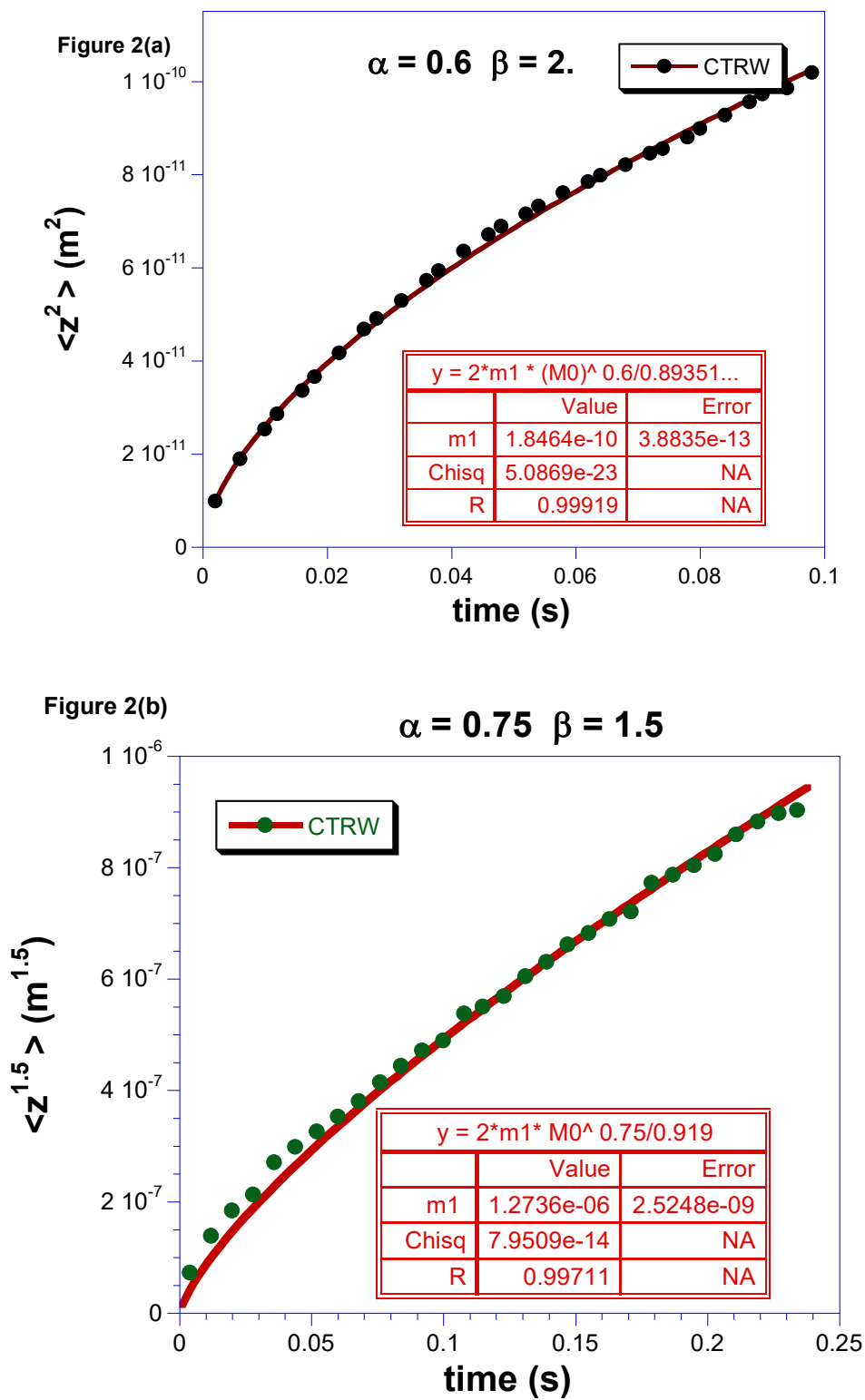

**Fig. 2** $\langle z^\beta(t) \rangle$ versus $t$ from the simulation: (a) $\alpha = 0.6, \beta = 2$, the fractional diffusion constant determined from the fitting is $D_f = 1.85 \times 10^{-10}$ m$^\beta$/s$^\alpha$, (b) $\alpha = 0.75, \beta = 1.5$, the fractional diffusion constant determined from the fitting is $D_f = 1.27 \times 10^{-6}$ m$^\beta$/s$^\alpha$.

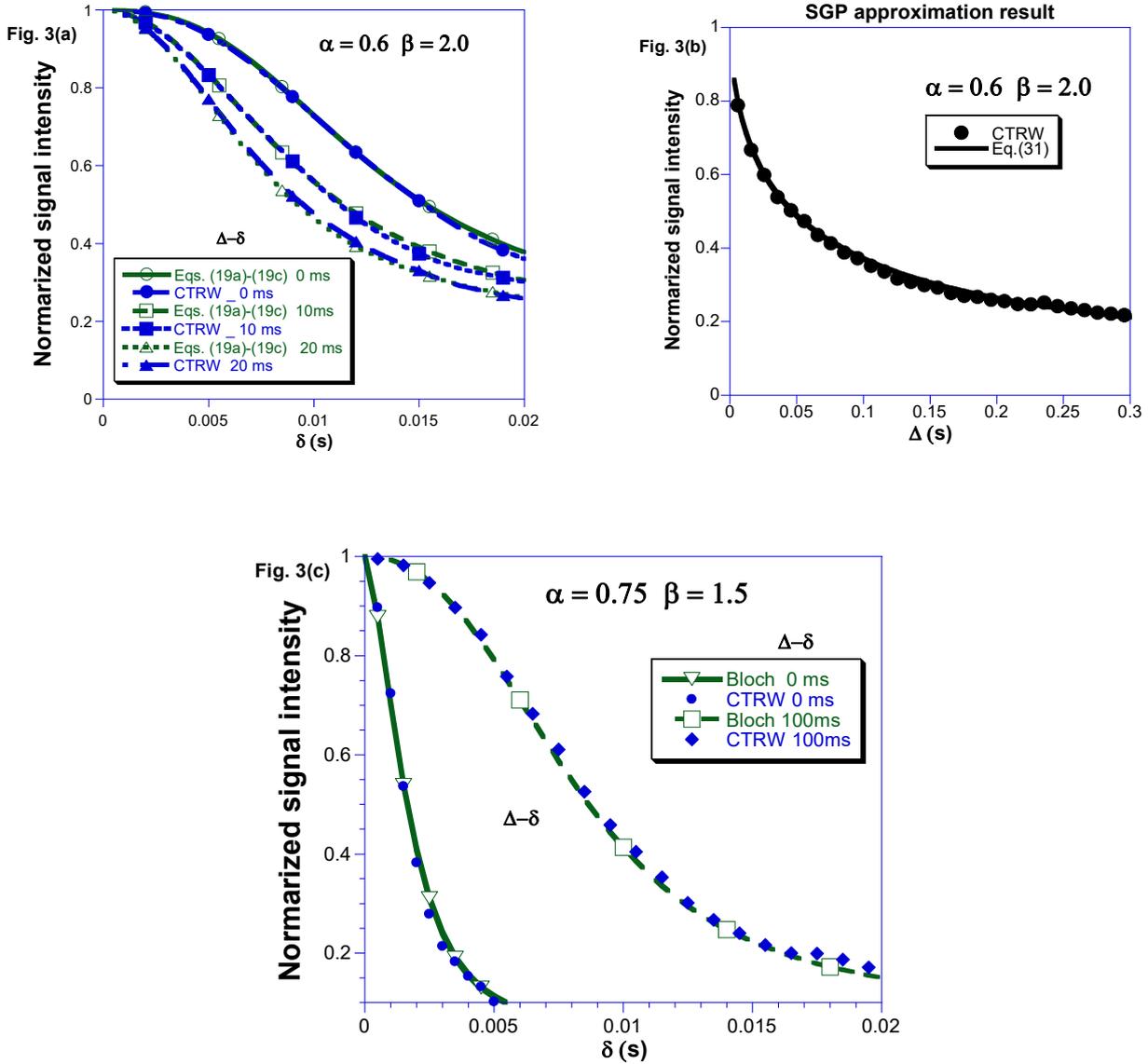

**Fig. 3** Comparison PFG signal attenuation from Eqs. (16a)-(16d) with that obtained from CTRW simulation: (a) $\alpha=0.6, \beta=2$ and $D_f = 1.85 \times 10^{-10}$ m$^\beta$/s$^\alpha$, finite gradient pulse width effect with $\Delta-\delta$ equaling 0 ms, 10 ms and 20 ms, $g$ equaling 0.1 $T/m$, (b) SGP approximation result, $\alpha=0.6, \beta=2$, $D_f = 1.85\times 10^{-10}$ m$^\beta$/s$^\alpha$, and $\gamma g\delta = 6\times 2.6751\times 10^4$, (c) $\alpha=0.75, \beta=1.5$ with $\Delta-\delta$ equaling 0 ms and $D_f = 1.15 \times 10^{-6}$ m$^\beta$/s$^\alpha$, and 100 ms $D_f = 1.27 \times 10^{-6}$ m$^\beta$/s$^\alpha$, $g$ equaling 0.05 $T/m$. The relaxation effect is neglected.

Figs. 4(a) and 4(b) compare the PFG signal attenuations obtained from the fractional derivative model with $T_2$ relaxation effect. Figs. 4(a) and 4(b) show $\Delta-\delta=0$ ms and $\Delta-\delta=20$ ms results, respectively. Other parameters used in Figs. 4(a) and 4(b) are $\alpha=0.6, \beta=2$, $D_f = 1.85 \times 10^{-10}$ m$^\beta$/s$^\alpha$, $g$ equaling 0.1 $T/m$, and $T_2$ equaling 50 ms. Figs. 4(a) and 4(b) indicate that the signal attenuation $S(t)$ including the relaxation effect based on Eq. (19e) is not proportional to $\exp(-t/T_2)$, while Eq. (19d) is proportional to $\exp(-t/T_2)$. When the signal attenuation resulted from the PFG fractional diffusion is large. The effective $T_2$ relaxation based on Eq. (19e) in PFG anomalous diffusion is slower than that based on Eq. (19d). The relaxation behavior in PFG fractional diffusion based on Eq. (19e) is significantly different from the normal diffusion where $S(t)$ affected by relaxation is proportional to $\exp(-t/T_2)$, which may be because Eq. (19e) includes the memory effect of the anomalous diffusion which affects the total signal attenuation resulting from the relaxation. In the PFG fractional diffusion, if at a local region, the relaxation could be viewed as the results of a small magnetization interacting with relaxation related Hamiltonian, then the relaxation may be relevant with PFG diffusion and Eq. (19e) maybe a choice as diffusion changes the local magnetization instantaneously. While it may be possible that the relaxation is not relevant to the attenuation of diffusion, then the general attenuation may obey Eq. (19d). Currently, it is unclear which one of these two equations, Eqs. (16d) and (16e), agrees with experimental results. It requires further experimental efforts to test these two equations.

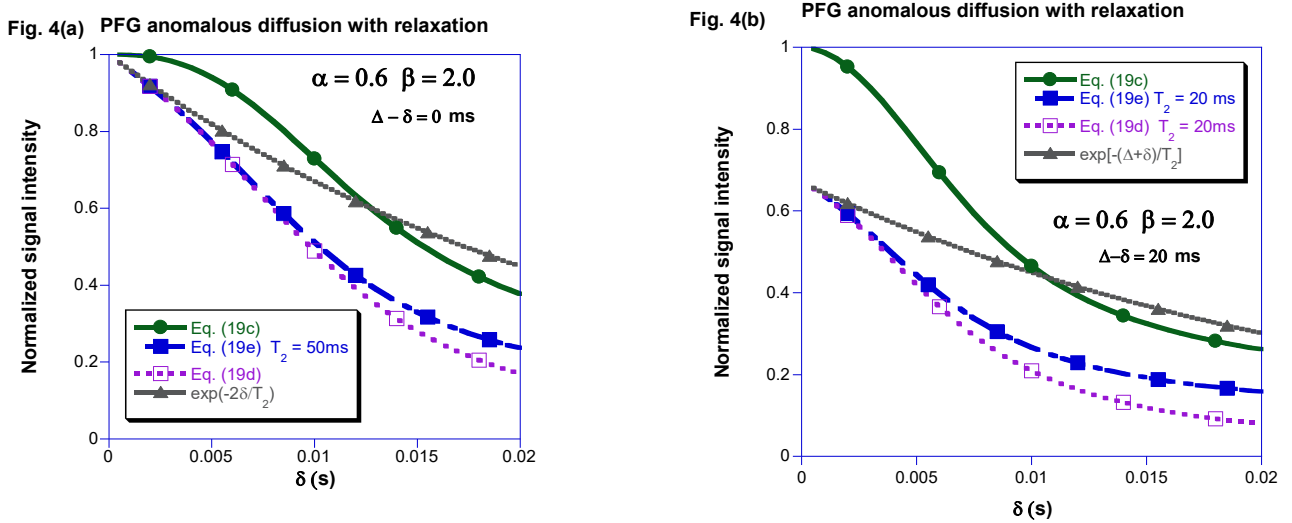

**Fig. 4** Comparison of Eqs. (16d) and (16e) that describe two different situations of Spin-spin relaxation effect on PFG anomalous diffusion, as described by the fractional derivative. $\alpha=0.6, \beta=2$, $D_f = 1.85 \times 10^{-10}$ m$^\beta$/s$^\alpha$, and $T_2$ is 50 ms are used: (a) $\Delta-\delta=0$ ms, (b) $\Delta-\delta=0$ ms. The signal attenuation $S(t)$ is proportional to $\exp(-\frac{t}{T_2})$ in Eq. (16d), but it is not proportional to Eq. (16e).

The numerical calculation by the Adomian decomposition method is simple for small signal attenuation. While at very large signal attenuation it may encounter overflow. The same numerical results can be obtained by an alternate method, the direct integration method. The agreement between these two methods can be seen in Fig. 5a. Compared to the Adomian decomposition method, the direct integration method has obvious advantages in numerical evaluation. It is not only fast, but also does not cause overflow. Additionally, based on Eq. (25), the direct integration provides a reliable way to calculate Mittag-Leffler function, $E_{\alpha,1}(-t^{\alpha})$, $t \in R$ because it does not cause overflow. The FORTRAN code for Mittag-Leffler function calculation can be obtained from the link https://github.com/GLin2017/Mittag-Leffler-function-calculated-by-Direct-Integration. The Mittag-Leffler function calculated by the direct integration method agrees with those calculated by other methods [45,46], which is demonstrated in Fig. 5b. The direct integration method provides a practical numerical evaluation method for analyzing PFG anomalous diffusion.

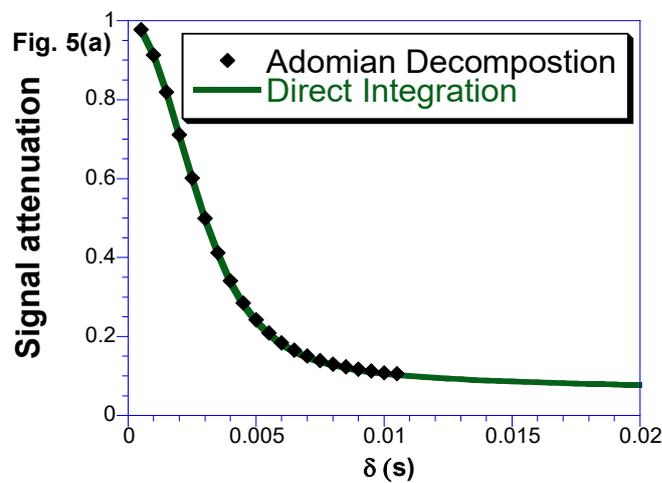

Fig. 5(a)

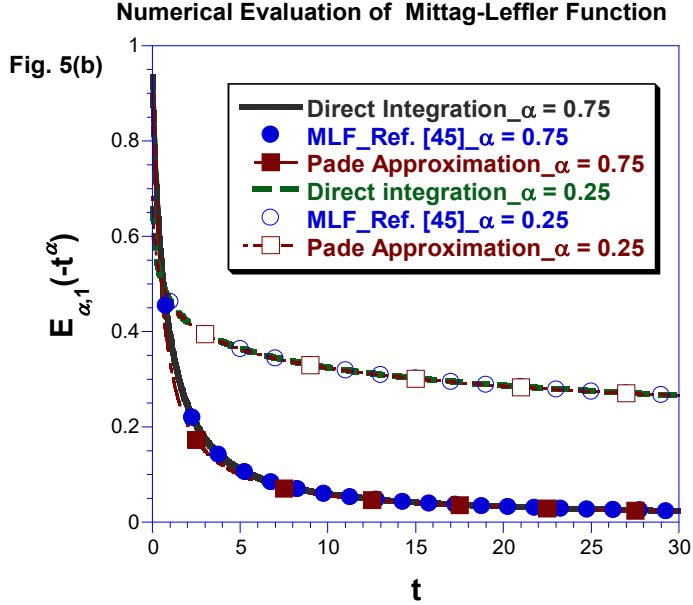

**Fig. 5** The numerical evaluation by the direct integration method (DIM): (a) the direct integration method agrees with the Adomian decomposition method in the numerical evaluation of PFG signal attenuation, the parameters used are $\alpha = 0.7$, $\beta = 2$, $D_f = 1.0 \times 10^{-9}$ m$^2$/s$^{0.7}$, $\Delta - \delta = 50$ ms and $g$ equaling 0.1 $T/m$, neglecting the relaxation effect in the calculation, (b) in calculation of the Mittag-Leffler function $E_{\alpha,1}(-t^\alpha)$, the direct integration method agrees with the method in Ref. [45] and the Pade approximation method in Ref. [46].

The modified-Bloch equation proposed in this paper is different from those reported Refs. [14] and [16]. The manner of combining the diffusion and precession is different from those used in those previously reported modified-Bloch equations [14,16]. In this paper, the different processes such as diffusion and precession are combined with the same time increment $dt$. The original properties of the contributions from linear or nonlinear processes remain unchanged at the instant of the combination, which is crucial in building modified-Bloch equations. The proposed modified-Bloch equation provides a fundamental equation for PFG anomalous diffusion in NMR and MRI research.

This paper focuses on the simple free anomalous diffusion in a homogeneous system. The anomalous diffusion can be non-symmetric diffusion, which can be described by the time-space fractional diffusion equation proposed in Ref. [21,22]. To investigate the non-symmetric diffusion, the space operator $\frac{\partial^\beta}{\partial |z|^\beta}$ need to be changed to a non-symmetric fractional space derivative such as Riesz-Feller space fractional derivative $_zD_\theta^\beta$, where the skewness $\theta$ affects the non-symmetric probability distribution function [21]. When $\theta = 0$, $_zD_\theta^\beta$ reduces to $\frac{\partial^\beta}{\partial |z|^\beta}$. However, this paper only focuses on the symmetric diffusion case because it is the most important starting point in PFG anomalous diffusion. Additionally, the real situation can be more complicated due to restricted diffusion [20,29,30], inhomogeneous distribution, nonlinear gradient field [26], anisotropic diffusion [39], etc.  Additional insights for those complicated PFG anomalous diffusion studies could be gained from other methods such as the instantaneous signal attenuation method, non-Gaussian approximation method, the effective phase shift diffusion equation method, observing the signal intensity at the origin method [47], SGP method [18,26], etc. It is worth mentioning that Refs. [48,49] could provide further insight into PFG diffusion based on Feynman-Kac theory.

## 5. Conclusion

This paper proposes a general routine to build the modified-Bloch equation for PFG anomalous diffusion. A fractional integral modified-Bloch equation was built, and its general solution was obtained. The major conclusions are summarized in the following:

1. The modified-Bloch equation is a combination of different processes such as diffusion, precession, and relaxation. These processes should keep their linear or nonlinear properties unchanged at the instant of the combination, which is critical criteria for building a modified-Bloch equation.
2. The general PFG signal attenuation expression is obtained. The expression includes the finite gradient pulse width effect, which is important for potential PFG anomalous diffusion applications such as these in the MRI.
3. The numerical evaluation can be done by the Adomian decomposition method or the direct integration method. These two methods give the identical results, but the direct integration method is much faster and does not cause overflow. Direct integration provides a convenient way to evaluate the Mittag-Leffler function type PFG signal attenuation.
4. The CTRW simulations agree with the theoretical results. The CTRW simulation provides a valuable method to analyze the PFG anomalous diffusion.
5. The spin-spin relaxation effect in PFG anomalous diffusion may deviate from $\exp(-t/T_2)$.

## Appendix A: Definition of the fractional derivative

The definition of the space fractional derivative [1,21-23] is given by

$$\frac{d^\beta}{d|z|^\beta} = -\frac{1}{2\cos\frac{\pi\alpha}{2}}\left[{_{-\infty}}D_z^\beta + {_z}D_\infty^\beta\right], \tag{A.1}$$

where

$$_{-\infty}D_z^\beta f(z) = \frac{1}{\Gamma(m-\beta)}\frac{d^m}{dz^m}\int_{-\infty}^z \frac{f(y)dy}{(z-y)^{\beta+1-m}}, \beta > 0, m-1 < \beta < m, \tag{A.2}$$

and

$$_zD_\infty^\beta f(z) = \frac{(-1)^m}{\Gamma(m-\beta)}\frac{d^m}{dz^m}\int_z^\infty \frac{f(y)dy}{(y-z)^{\beta+1-m}}, \beta > 0, m-1 < \beta < m. \tag{A.3}$$

**Appendix B: Derive PFG signal attenuation equation from modified-Bloch Equation**

In section 2.1, the PFG signal attenuation equation was obtained from the fractional modified-Bloch equation (11) by substituting $M_{xy}(\mathbf{z},\tau) = S(\tau)\exp(-i\mathbf{K}(\tau)\cdot\mathbf{z})$ into Eq. (11) and using the origin properties: $\mathbf{z} = 0$ and $\exp(-i\mathbf{K}(t)\cdot\mathbf{z}) = 1$.

There are other ways to derive the PFG signal attenuation equation (12) from the fractional modified-Bloch equation (11).

The modified-Bloch equation Eq. (11) is built based on the simultaneous equations (3) and (9). According to Eq. (3), the gradient field induced precession changes the magnetization phase by $\exp(-i\gamma\mathbf{g}(t)\cdot\mathbf{z}dt)$ during small time interval $dt$

$$M_{xy,\tau}(\mathbf{z},t) - i\gamma\mathbf{g}(\tau)\cdot\mathbf{z}M_{xy,\tau}(\mathbf{z},t)d\tau = \exp[-i\gamma\mathbf{g}(\tau)\cdot\mathbf{z}d\tau]M_{xy,\tau}(\mathbf{z},t) \ . \tag{B.1}$$

Let us calculate $M_{xy,\tau}(\mathbf{z},t)$ based on Eq. (11) step by step starting from $d\tau$ to time $t$ with interval $d\tau$. For time $1d\tau$,

$$M_{xy,d\tau}(\mathbf{z},t) = \exp[-i\gamma\mathbf{g}(d\tau)\cdot\mathbf{z}d\tau]\left\{\sum_{k=0}^{m-1} M_{xy}^{(k)}(\mathbf{z},0^+)\frac{t^k}{k!} + \left[\frac{(t-0)^{\alpha-1}}{\Gamma(\alpha)}D_f\frac{\partial^\beta}{\partial|z|^\beta}M_{xy}(\mathbf{z},0)\right]d\tau\right\}. \tag{B.2}$$

Note in the above discrete calculation, $M_{xy,d\tau}(\mathbf{z},t)$ is obtained based on $M_{xy,d\tau-d\tau}(\mathbf{z},t)$ and $M_{xy}(\mathbf{z},d\tau-d\tau) = M_{xy}(\mathbf{z},0)$. In general, $M_{xy,id\tau}(\mathbf{z},t)$ of the $i^{th}$ step can be calculated based on $M_{xy}(\mathbf{z},(i-1)d\tau)$ of the $(i-1)^{th}$ step. For time $2d\tau$,

$$M_{xy,2d\tau}(\mathbf{z},t) = \exp[-i\gamma\mathbf{g}(2d\tau)\cdot\mathbf{z}d\tau]\left\{M_{xy,1d\tau}(\mathbf{z},t) + \left[\frac{(t-d\tau)^{\alpha-1}}{\Gamma(\alpha)}D_f\frac{\partial^\beta}{\partial|z|^\beta}M_{xy}(\mathbf{z},d\tau)\right]d\tau\right\},$$

Substituted expression (B.2), $M_{xy,d\tau}(\mathbf{z},t)$ into $M_{xy,2d\tau}(\mathbf{z},t)$, we get

$$\begin{aligned}
M_{xy,2d\tau}(\mathbf{z},t) &= \exp[-i\gamma\mathbf{g}(2d\tau)\cdot\mathbf{z}d\tau]\bigg(\exp[-i\gamma\mathbf{g}(d\tau)\cdot\mathbf{z}d\tau]\bigg\{\sum_{k=0}^{m-1} M_{xy}^{(k)}(\mathbf{z},0^+)\frac{t^k}{k!} + \\
&\quad \left[\frac{(t-0)^{\alpha-1}}{\Gamma(\alpha)}D_f\frac{\partial^\beta}{\partial|z|^\beta}M_{xy}(\mathbf{z},0)\right]d\tau\bigg\} + \left[\frac{(t-d\tau)^{\alpha-1}}{\Gamma(\alpha)}D_f\frac{\partial^\beta}{\partial|z|^\beta}M_{xy}(\mathbf{z},d\tau)\right]d\tau\bigg) \\
&= \exp[-i\gamma\mathbf{g}(2d\tau)\cdot\mathbf{z}d\tau]\exp[-i\gamma\mathbf{g}(d\tau)\cdot\mathbf{z}d\tau]\sum_{k=0}^{m-1} M_{xy}^{(k)}(\mathbf{z},0^+)\frac{t^k}{k!} + \\
&\quad \exp[-i\gamma\mathbf{g}(2d\tau)\cdot\mathbf{z}d\tau]\exp[-i\gamma\mathbf{g}(d\tau)\cdot\mathbf{z}d\tau]\left\{\left[\frac{(t-0)^{\alpha-1}}{\Gamma(\alpha)}D_f\frac{\partial^\beta}{\partial|z|^\beta}M_{xy}(\mathbf{z},0)\right]d\tau\right\} + \\
&\quad \exp[-i\gamma\mathbf{g}(2d\tau)\cdot\mathbf{z}d\tau]\left[\frac{(t-d\tau)^{\alpha-1}}{\Gamma(\alpha)}D_f\frac{\partial^\beta}{\partial|z|^\beta}M_{xy}(\mathbf{z},d\tau)\right]d\tau.
\end{aligned}$$

(B.3)

Similarly,

$$M_{xy,3d\tau}(\mathbf{z},t) = \exp[-i\gamma\mathbf{g}(3d\tau)\cdot\mathbf{z}d\tau]\left\{M_{xy,2d\tau}(\mathbf{z},t) + \left[\frac{(t-2d\tau)^{\alpha-1}}{\Gamma(\alpha)}D_f\frac{\partial^\beta}{\partial|z|^\beta}M_{xy}(\mathbf{z},2d\tau)\right]d\tau\right\}$$

$$= \exp[-\sum_{l=1}^{3}i\gamma\mathbf{g}(ld\tau)\cdot\mathbf{z}d\tau]\sum_{k=0}^{m-1}M_{xy}^{(k)}(\mathbf{z},0^+)\frac{t^k}{k!} +$$

$$\exp[-\sum_{l=1}^{3}i\gamma\mathbf{g}(ld\tau)\cdot\mathbf{z}d\tau]\left\{\left[\frac{(t-0)^{\alpha-1}}{\Gamma(\alpha)}D_f\frac{\partial^\beta}{\partial|z|^\beta}M_{xy}(\mathbf{z},0)\right]d\tau\right\}$$

$$+ \exp[-\sum_{l=2}^{3}i\gamma\mathbf{g}(ld\tau)\cdot\mathbf{z}d\tau]\left[\frac{(t-d\tau)^{\alpha-1}}{\Gamma(\alpha)}D_f\frac{\partial^\beta}{\partial|z|^\beta}M_{xy}(\mathbf{z},d\tau)\right]d\tau$$

$$+ \exp[-i\gamma\mathbf{g}(3d\tau)\cdot\mathbf{z}d\tau]\left[\frac{(t-2d\tau)^{\alpha-1}}{\Gamma(\alpha)}D_f\frac{\partial^\beta}{\partial|z|^\beta}M_{xy}(\mathbf{z},2d\tau)\right]d\tau,$$

(B.4)

and

$$M_{xy,nd\tau}(\mathbf{z},t) = \exp[-\sum_{l=1}^{n}i\gamma\mathbf{g}(ld\tau)\cdot\mathbf{z}d\tau]\sum_{k=0}^{m-1}M_{xy}^{(k)}(\mathbf{z},0^+)\frac{t^k}{k!}$$
$$+ \sum_{i=1}^{n}\left(\exp[-\sum_{l=i}^{n}i\gamma\mathbf{g}(ld\tau)\cdot\mathbf{z}d\tau]\left[\frac{[t-(i-1)d\tau]^{\alpha-1}}{\Gamma(\alpha)}D_f\frac{\partial^\beta}{\partial|z|^\beta}M_{xy}(\mathbf{z},(i-1)d\tau)\right]d\tau\right),$$

(B.5)

which can be rewritten in an integragtion form as

$$M_{xy,nd\tau}(\mathbf{z},t) = \sum_{k=0}^{m-1}M_{xy}^{(k)}(\mathbf{z},0^+)\frac{t^k}{k!}\exp\left(-\int_0^{nd\tau}i\gamma\mathbf{g}(\tau)\cdot\mathbf{z}d\tau\right) +$$

$$\int_0^{nd\tau}\left[\exp\left(-\int_\tau^{nd\tau}i\gamma\mathbf{g}(\tau')\cdot\mathbf{z}d\tau'\right)\frac{(t-\tau)^{\alpha-1}}{\Gamma(\alpha)}D_f\frac{\partial^\beta}{\partial|z|^\beta}M_{xy}(\mathbf{z},\tau)\right]d\tau,$$

(B.6)

where $(i-1)d\tau$ is replaced with $id\tau$, and $\frac{\partial^\beta}{\partial|z|^\beta}M_{xy}(\mathbf{z},(i-1)d\tau)$ is replaced by $\frac{\partial^\beta}{\partial|z|^\beta}M_{xy}(\mathbf{z},id\tau)$. These replacements are reasonable because their differences are negligible with $d\tau$ can be arbitrarily set as an infinitely small interval. When $nd\tau = t$,

$$M_{xy}(\mathbf{z},t) = \sum_{k=0}^{m-1}M_{xy}^{(k)}(\mathbf{z},0^+)\frac{t^k}{k!}\exp\left(-\int_0^t i\gamma\mathbf{g}(\tau')\cdot\mathbf{z}d\tau'\right) +$$

$$\int_0^t\left[\exp\left(-\int_\tau^t i\gamma\mathbf{g}(\tau')\cdot\mathbf{z}d\tau'\right)\frac{(t-\tau)^{\alpha-1}}{\Gamma(\alpha)}D_f\frac{\partial^\beta}{\partial|z|^\beta}M_{xy}(\mathbf{z},\tau)\right]d\tau.$$

(B.7)

Substituting $M_{xy}(\mathbf{z},\tau) = S(\tau)\exp(-i\mathbf{K}(\tau)\cdot\mathbf{z})$ into Eq. (B.7), we have

$$S(t)\exp(-i\mathbf{K}(t)\cdot\mathbf{z}) = \sum_{k=0}^{m-1} M_{xy}^{(k)}(\mathbf{z},0^+)\frac{t^k}{k!}\exp(-i\mathbf{K}(t)\cdot\mathbf{z}) +$$
$$\int_0^t \left[\exp\left(-\int_\tau^t i\gamma\mathbf{g}(\tau')\cdot\mathbf{z}d\tau'\right)\frac{(t-\tau)^{\alpha-1}}{\Gamma(\alpha)}D_f|K(\tau)|^\beta S(t)\exp(-i\mathbf{K}(\tau)\cdot\mathbf{z})\right]d\tau.$$

(B.8)

After eliminating the term $\exp(-i\mathbf{K}(t)\cdot\mathbf{z})$ from both sides of Eq. (B.8), we get

$$S(t) = \sum_{k=0}^{m-1} M_{xy}^{(k)}(\mathbf{z},0^+)\frac{t^k}{k!} + \int_0^t \left[\frac{(t-\tau)^{\alpha-1}}{\Gamma(\alpha)}D_f|K(t)|^\beta \frac{\partial^\beta}{\partial|z|^\beta}S(\tau)\right]d\tau.$$

(B.9)

Eq. (B.9) is the same as Eq. (12)

Additionally, Eq. (B.7) can be linearly transformed to

$$\exp(i\phi)M_{xy}(\mathbf{z},t) = \sum_{k=0}^{m-1} M_{xy}^{(k)}(\mathbf{z},0^+)\frac{t^k}{k!}\exp(i\phi)\exp\left(-\int_0^t i\gamma\mathbf{g}(\tau')\cdot\mathbf{z}d\tau'\right) +$$
$$\int_0^t \left[\exp(i\phi)\exp\left(-\int_\tau^t i\gamma\mathbf{g}(\tau')\cdot\mathbf{z}d\tau'\right)\frac{(t-\tau)^{\alpha-1}}{\Gamma(\alpha)}D_f\frac{\partial^\beta}{\partial|z|^\beta}\exp(-i\phi)\exp(i\phi)M_{xy}(\mathbf{z},\tau)\right]d\tau,$$

(B.10)

where $\phi$ is the rotation angle. When $\exp(i\phi) = \exp(i\mathbf{K}(t)\cdot\mathbf{z})$, we get

$$\exp(i\mathbf{K}(t)\cdot\mathbf{z})M_{xy}(\mathbf{z},t) = \sum_{k=0}^{m-1} M_{xy}^{(k)}(\mathbf{z},0^+)\frac{t^k}{k!} + \int_0^t \left[\exp(i\mathbf{K}(\tau)\cdot\mathbf{z})\frac{(t-\tau)^{\alpha-1}}{\Gamma(\alpha)}D_f\frac{\partial^\beta}{\partial|z|^\beta}M_{xy}(\mathbf{z},\tau)\right]d\tau,$$

(B.11)

which is equivalent to

$$_tD_*^\alpha\left[\exp(i\mathbf{K}(t)\cdot\mathbf{z})M_{xy}(\mathbf{z},t)\right] = \exp(i\mathbf{K}(t)\cdot\mathbf{z})D_f\frac{\partial^\beta}{\partial|z|^\beta}M_{xy}(\mathbf{z},t).$$

(B.12)

For a single pulse attenuation as mentioned in Section 4, $K(t) = \gamma\mathbf{g}\cdot\mathbf{z}$. Eq. (B.12) agrees the results from modified-Bloch equation proposed in Ref. [16]. Substituting $M_{xy}(\mathbf{z},\tau) = S(\tau)\exp(-i\mathbf{K}(\tau)\cdot\mathbf{z})$ into Eq. (B.12), we get Eq. (12) again.

However, in an inhomogeneous system or restricted diffusion systems, the real magnetization phase does not linearly depend on the position $z$, which thus cannot be canceled by the linear transformation by multiplying with $\exp(i\mathbf{K}(t)\cdot\mathbf{z})$. In these complicated systems, the magnetization can be calculated from Eq. (11) step by step.

**Appendix C: Obtain the magnetization directly from the modified-Bloch equation**

From Eq. (11), we can calculate $M_{xy}(\mathbf{z},d\tau), M_{xy}(\mathbf{z},2d\tau), M_{xy}(\mathbf{z},3d\tau),\ldots M_{xy}(\mathbf{z},t)$ step by step starting from time $d\tau$ to time $t = nd\tau$ with equal time increment $d\tau$. For a homogeneous spin system, $M_{xy}(\mathbf{z},0) = \sum_{k=0}^{m-1} M_{xy}^{(k)}(\mathbf{z},0^+)\frac{t^k}{k!} = 1$, and $\frac{\partial^\beta}{\partial|z|^\beta}M_{xy}(\mathbf{z},0) = 0$, we thus have

$$M_{xy}(\mathbf{z}, d\tau) = M_{xy,d\tau}(\mathbf{z}, d\tau)$$

$$= \exp[-i\gamma \mathbf{g}(d\tau) \cdot \mathbf{z} d\tau] \left\{ \sum_{k=0}^{m-1} M_{xy}^{(k)}(\mathbf{z}, 0^+) \frac{t^k}{k!} + \left[ \frac{(t-0)^{\alpha-1}}{\Gamma(\alpha)} D_f \frac{\partial^\beta}{\partial |z|^\beta} M_{xy}(\mathbf{z}, 0) \right] d\tau \right\}$$

$$= \exp[-i\gamma \mathbf{g}(d\tau) \cdot \mathbf{z} d\tau] S(d\tau),$$

(C.1)

where $S(d\tau) = \sum_{k=0}^{m-1} M_{xy}^{(k)}(\mathbf{z}, 0^+) \frac{t^k}{k!} = 1$. The calculation of $M_{xy}(\mathbf{z}, 2d\tau)$ needs two discrete steps, one for $M_{xy,d\tau}(\mathbf{z}, 2d\tau)$ and the other for $M_{xy}(\mathbf{z}, 2d\tau)$. $M_{xy,d\tau}(\mathbf{z}, 2d\tau)$ can be calculated as

$$M_{xy,d\tau}(\mathbf{z}, 2d\tau) = \exp[-i\gamma \mathbf{g}(d\tau) \cdot \mathbf{z} d\tau] \left\{ \sum_{k=0}^{m-1} M_{xy}^{(k)}(\mathbf{z}, 0^+) \frac{t^k}{k!} + \left[ \frac{(2d\tau-0)^{\alpha-1}}{\Gamma(\alpha)} D_f \frac{\partial^\beta}{\partial |z|^\beta} M_{xy}(\mathbf{z}, 0) \right] d\tau \right\}$$

$$= \exp[-i\gamma \mathbf{g}(d\tau) \cdot \mathbf{z} d\tau],$$

$$M_{xy}(\mathbf{z}, 2d\tau) = \exp[-i\gamma \mathbf{g}(2d\tau) \cdot \mathbf{z} d\tau] \left\{ M_{xy,d\tau}(\mathbf{z}, 2d\tau) + \left[ \frac{(2d\tau-d\tau)^{\alpha-1}}{\Gamma(\alpha)} D_f \frac{\partial^\beta}{\partial |z|^\beta} M_{xy}(\mathbf{z}, d\tau) \right] d\tau \right\}$$

$$= \exp[-i\gamma \mathbf{g}(2d\tau) \cdot \mathbf{z} d\tau] \exp[-i\gamma \mathbf{g}(d\tau) \cdot \mathbf{z} d\tau] -$$

$$\exp[-i\gamma \mathbf{g}(2d\tau) \cdot \mathbf{z} d\tau] \exp[-i\gamma \mathbf{g}(d\tau) \cdot \mathbf{z} d\tau] \left\{ \left[ \frac{(2d\tau-d\tau)^{\alpha-1}}{\Gamma(\alpha)} D_f |\gamma \mathbf{g}(d\tau) d\tau|^\beta S(d\tau) \right] d\tau \right\}$$

$$= \exp[-i\gamma \mathbf{g}(2d\tau) \cdot \mathbf{z} d\tau] \exp[-i\gamma \mathbf{g}(d\tau) \cdot \mathbf{z} d\tau] \left\{ 1 - \left[ \frac{(2d\tau-d\tau)^{\alpha-1}}{\Gamma(\alpha)} D_f |\gamma \mathbf{g}(d\tau) d\tau|^\beta S(d\tau) \right] d\tau \right\}$$

$$= \exp[-i\gamma \mathbf{g}(2d\tau) \cdot \mathbf{z} d\tau] \exp[-i\gamma \mathbf{g}(d\tau) \cdot \mathbf{z} d\tau] S(2d\tau),$$

(C.2)

where

$$S(2d\tau) = 1 - \left[ \frac{(2d\tau-d\tau)^{\alpha-1}}{\Gamma(\alpha)} D_f |\gamma \mathbf{g}(d\tau) d\tau|^\beta S(d\tau) \right] d\tau.$$

(C.3)

Similarly, we can get

$$M_{xy,d\tau}(\mathbf{z}, 3d\tau) = \exp[-i\gamma \mathbf{g}(d\tau) \cdot \mathbf{z} d\tau] \left\{ 1 + \left[ \frac{(3d\tau-d\tau)^{\alpha-1}}{\Gamma(\alpha)} D_f \frac{\partial^\beta}{\partial |z|^\beta} M_{xy}(\mathbf{z}, 0) \right] d\tau \right\}$$

$$= \exp[-i\gamma \mathbf{g}(d\tau) \cdot \mathbf{z} d\tau],$$

$$M_{xy,2d\tau}(\mathbf{z}, 3d\tau) = \exp[-i\gamma \mathbf{g}(2d\tau) \cdot \mathbf{z} d\tau] \left\{ M_{xy,d\tau}(\mathbf{z}, 3d\tau) + \left[ \frac{(3d\tau-d\tau)^{\alpha-1}}{\Gamma(\alpha)} D_f \frac{\partial^\beta}{\partial |z|^\beta} M_{xy}(\mathbf{z}, d\tau) \right] d\tau \right\}$$

$$= \exp[-i\gamma \mathbf{g}(2d\tau) \cdot \mathbf{z} d\tau] \exp[-i\gamma \mathbf{g}(d\tau) \cdot \mathbf{z} d\tau] \left\{ 1 - \left[ \frac{(3d\tau-d\tau)^{\alpha-1}}{\Gamma(\alpha)} D_f |\gamma \mathbf{g}(d\tau) d\tau|^\beta S(d\tau) \right] d\tau \right\},$$

$$M_{xy}(\mathbf{z}, 3d\tau) = M_{xy,3d\tau}(\mathbf{z}, 3d\tau)$$

$$= \exp[-i\gamma\mathbf{g}(3d\tau) \cdot \mathbf{z}d\tau]\left\{M_{xy,2d\tau}(\mathbf{z}, 3d\tau) + \left[\frac{(3d\tau - 2d\tau)^{\alpha-1}}{\Gamma(\alpha)}D_f \frac{\partial^\beta}{\partial|z|^\beta}M_{xy}(\mathbf{z}, 2d\tau)\right]d\tau\right\}$$

$$= \exp[-\sum_{l=1}^{3} i\gamma\mathbf{g}(ld\tau) \cdot \mathbf{z}d\tau]\left\{1 - \left[\frac{(3d\tau - d\tau)^{\alpha-1}}{\Gamma(\alpha)}D_f|\gamma\mathbf{g}(d\tau)d\tau|^\beta S(d\tau)\right]d\tau\right\}$$

$$- \exp[-\sum_{l=1}^{3} i\gamma\mathbf{g}(ld\tau) \cdot \mathbf{z}d\tau]\frac{(3d\tau - 2d\tau)^{\alpha-1}}{\Gamma(\alpha)}D_f|\gamma\mathbf{g}(d\tau)d\tau + \gamma\mathbf{g}(2d\tau)d\tau|^\beta S(2d\tau)d\tau$$

$$= \exp[-\sum_{l=1}^{3} i\gamma\mathbf{g}(ld\tau) \cdot \mathbf{z}d\tau]\left\{1 - \left[\frac{(3d\tau - d\tau)^{\alpha-1}}{\Gamma(\alpha)}D_f|\gamma\mathbf{g}(d\tau)d\tau|^\beta S(d\tau)\right]d\tau - \right.$$

$$\left. \left[\frac{(3d\tau - 2d\tau)^{\alpha-1}}{\Gamma(\alpha)}D_f|\gamma\mathbf{g}(d\tau)d\tau + \gamma\mathbf{g}(2d\tau)d\tau|^\beta S(2d\tau)\right]d\tau\right\}$$

$$= \exp[-\sum_{l=1}^{3} i\gamma\mathbf{g}(ld\tau) \cdot \mathbf{z}d\tau]S(3d\tau),$$

(C.4)

where

$$S(3d\tau) = 1 - \left[\frac{(3d\tau - d\tau)^{\alpha-1}}{\Gamma(\alpha)}D_f|\gamma\mathbf{g}(d\tau)d\tau|^\beta S(d\tau)\right]d\tau -$$

$$\left[\frac{(3d\tau - 2d\tau)^{\alpha-1}}{\Gamma(\alpha)}D_f|\gamma\mathbf{g}(d\tau)d\tau + \gamma\mathbf{g}(2d\tau)d\tau|^\beta S(2d\tau)\right]d\tau;$$

(C.5)

at $t = nd\tau$,

$$M_{xy}(\mathbf{z}, t) = M_{xy,3d\tau}(\mathbf{z}, nd\tau)$$

$$= \exp[-i\gamma\mathbf{g}(nd\tau) \cdot \mathbf{z}d\tau]\left\{M_{xy,(n-1)d\tau}(\mathbf{z}, nd\tau) + \left[\frac{(nd\tau - (n-1)d\tau)^{\alpha-1}}{\Gamma(\alpha)}D_f \frac{\partial^\beta}{\partial|z|^\beta}M_{xy}(\mathbf{z}, (n-1)d\tau)\right]d\tau\right\}$$

$$= \exp[-\sum_{l=1}^{n} i\gamma\mathbf{g}(ld\tau) \cdot \mathbf{z}d\tau]\left\{1 - \left[\frac{(nd\tau - d\tau)^{\alpha-1}}{\Gamma(\alpha)}D_f|\gamma\mathbf{g}(d\tau)d\tau|^\beta S(d\tau)\right]d\tau - \right.$$

$$\left[\frac{(nd\tau - 2d\tau)^{\alpha-1}}{\Gamma(\alpha)}D_f|\gamma\mathbf{g}(d\tau)d\tau + \gamma\mathbf{g}(2d\tau)d\tau|^\beta S(2d\tau)\right]d\tau \ldots -$$

$$\left.\left[\frac{(nd\tau - (n-1)d\tau)^{\alpha-1}}{\Gamma(\alpha)}D_f\left|\sum_{j=1}^{n-1}\gamma\mathbf{g}(jd\tau)d\tau\right|^\beta S((n-1)d\tau)\right]d\tau\right\}$$

$$= \exp\left\{-\left[\int_0^t i\gamma\mathbf{g}(\tau)d\tau\right] \cdot \mathbf{z}\right\}\left\{1 - \int_0^t \frac{(t-\tau)^{\alpha-1}}{\Gamma(\alpha)}D_f\left|\int_0^\tau i\gamma\mathbf{g}(\tau')d\tau'\right|^\beta S(\tau)d\tau\right\}$$

$$= \exp[-\mathbf{K}(t) \cdot \mathbf{z}]\left\{1 - \int_0^t \frac{(t-\tau)^{\alpha-1}}{\Gamma(\alpha)}D_f|\mathbf{K}(\tau)|^\beta S(\tau)d\tau\right\}$$

$$= \exp[-\mathbf{K}(t) \cdot \mathbf{z}]S(t),$$

(C.6)

where $\mathbf{K}(t) = \int_0^t i\gamma \mathbf{g}(\tau)d\tau$ and

$$S(t) = \left\{1 - \int_0^t \frac{(t-\tau)^{\alpha-1}}{\Gamma(\alpha)} D_f |\mathbf{K}(\tau)|^\beta S(\tau)d\tau\right\}. \tag{C.7}$$

Eq. (C.6) gives the magnetization of homogenous spin system, which includes the phase $\exp[-\mathbf{K}(t) \cdot \mathbf{z}]$, and the signal amplitude $S(t)$ described by Eq. (C.7). Eq. (C.7) is the same as Eq. (12). Only the initial condition $\sum_{k=0}^{m-1} M_{xy}^{(k)}(\mathbf{z}, 0^+) \frac{t^k}{k!} = 1$ is used in the derivation (calculation). It is plausible that the whole magnetization expression including phase and amplitude can be directly obtained from the modified-Bloch equation Eq. (11). To the best of my knowledge, no other reported methods based on the fractional derivative have been able to achieve this. The above method could be used to obtain magnetization in complex diffusion system such as restricted diffusion in the future.

## Acknowledgements

The linguistic help from Thomas Caywood, Christian Fallen from writing center in Clark University, and Amoy Lin is acknowledged.

## References

[1] W. Wyss, J. Math. Phys. 27 (1986) 2782-2785.

[2] R. Metzler, J. Klafter, The random walk's guide to anomalous diffusion: a fractional dynamics approach, Phys. Rep. 339 (2000) 1-77.

[3] I.M. Sokolov, Models of anomalous diffusion in crowded environments, Soft Matter 8 (2012) 9043-9052.

[4] E.L. Hahn, Spin echoes, Phys. Rev. 80 (1950) 580-594.

[5] D. W. McCall, D. C. Douglass, E. W. Anderson, Ber. Bunsenges. Phys. Chem. 67 (1963) 336-340.

[6] E. O. Stejskal, J. E. Tanner, J. Chem. Phys. 42 (1965) 288-292, doi: 10.1063/1.1695690.

[7] J. Kärger, H. Pfeifer, G. Vojta, Time correlation during anomalous diffusion in fractal systems and signal attenuation in NMR field-gradient spectroscopy, Phys. Rev. A 37 (11) (1988) 4514-4517.

[8] N. Fatkullin, R. Kimmich, Phys. Rev. E 52 (1995) 3273-3276.

[9] R. A. Damion, K.J. Packer, Predictions for Pulsed-Field-Gradient NMR Experiments of Diffusion in Fractal Spaces, Proceedings: Mathematical, Physical and Engineering Sciences, 453 (1997) 205-211.

[10] T. Zavada, N. Südland, R. Kimmich, T. F. Nonnenmacher, Phys. Rev. E 60 (1999) 1292.

[11] K.M. Bennett, K.M. Schmainda, R.T. Bennett, D.B. Rowe, H. Lu, J.S. Hyde, Characterization of continuously distributed cortical water diffusion rates with a stretched-exponential model, Magn. Reson. Med. 50 (2003) 727-734.

[12] K.M. Bennett, J.S. Hyde, K.M. Schmainda, Water diffusion heterogeneity index in the human brain is insensitive to the orientation of applied magnetic field gradients, Magn. Reson. Med. 56 (2006) 235-239.

[13] E. Özarslan, P.J. Basser, T.M. Shepherd, P.E. Thelwall, B.C. Vemuri, S.J. Blackband, Observation of anomalous diffusion in excised tissue by characterizing the diffusion-time dependence of the MR signal, J. Magn. Reson. 183 (2006) 315.

[14] R.L. Magin, O. Abdullah, D. Baleanu, X.J. Zhou, Anomalous diffusion expressed through fractional order differential operators in the Bloch–Torrey equation, J. Magn. Reson. 190 (2008) 255-270.

[15] M. Palombo, A. Gabrielli, S.D. Santis, C. Cametti, G. Ruocco, S. Capuani, Spatio-temporal anomalous diffusion in heterogeneous media by nuclear magnetic resonance, J. Chem. Phys. 135 (2011) 034504.

[16] A. Hanyga, M. Seredyńska, Anisotropy in high-resolution diffusion-weighted MRI and anomalous diffusion, J. Magn. Reson. 220 (2012) 85-93.

[17] M. Palombo, A. Gabrielli, S. De Santis, C. Cametti, G. Ruocco, S. Capuani, J. Chem. Phys. 135 (2014) 034504.

[18] G. Lin, An effective phase shift diffusion equation method for analysis of PFG normal and fractional diffusions, J. Magn. Reson. 259 (2015) 232-240.

[19] G. Lin, Analyzing signal attenuation in PFG anomalous diffusion via a non-Gaussian phase distribution approximation approach by fractional derivatives, J. Chem. Phys. 145 (2016) 194202.

[20] G. Lin, S. Zheng, X. Liao, Signal attenuation of PFG restricted anomalous diffusions in plate, sphere, and cylinder, J. Magn. Reson. 272 (2016) 25-36.

[21] F. Mainardi, Yu. Luchko, G. Pagnini, The fundamental solution of the space-time-fractional diffusion diffusion equation, Fract. Calc. Appl. Anal. 4 (2001) 153-192.

[22] R. Gorenflo, F. Mainardi, Fractional Diffusion Processes: Probability Distributions and Continuous Time Random Walk, Springer Lecture Notes in Physics, No 621, Berlin, 2003, pp. 148–166.

[23] R. Balescu, V-Langevin equations, continuous time random walks and fractional diffusion, Chaos Soliton Fract. 34 (2007) 62-80.

[24] H. C. Torrey, Bloch Equations with Diffusion Terms, Phys. Rev. 104 (3) (1956) 563-565.

[25] G. Germano, M. Politi, E. Scalas, R. L. Schilling, Phys. Rev. E 79 (2009) 066102.

[26] G. Lin, Instantaneous signal attenuation method for analysis of PFG fractional Diffusions, J. Magn. Reson. 269 (2016) 36-49.

[27] Cicerone, M. T. Wagner, P. A. Ediger, M. D. Translational Diffusion on Heterogeneous Lattices: A Model for Dynamics in Glass Forming Materials. J. Phys. Chem. B 101 (1997) 8727.

[28] G. Lin, J. Zhang, H. Cao, A. A. Jones, J. Phys. Chem. B 107 (2003) 6179.

[29] P. Callaghan, Translational Dynamics and Magnetic Resonance: Principles of Pulsed Gradient Spin Echo NMR, Oxford University Press, 2011.

[30] W. S. Price, NMR Studies of Translational Motion, Cambridge University Press, Cambridge, UK, 2009.

[31] R. Kimmich, NMR: Tomography, Diffusometry, Relaxometry, Springer-Verlag, Heidelberg, 1997.

[32] N. Bloembergen, E. M. Purcell, R. V. Pound, Relaxation Effects in Nuclear Magnetic Resonance Absorption, Phys. Rev. 73 (7) (1948) 679-746.

[33] F. Bloch, Nuclear Induction, Phys. Rev. 70 (1946) 460-474.

[34] R. C. Mittal, R. Nigam, Solution of fractional integro-differential equations by Adomian decomposition method, Int. J. of Appl. Math. and Mech. 4 (2) (2008) 87-94.

[35] J. Klafter, I.M. Sokolov, First Step in Random Walks. From Tools to Applications, Oxford University Press, New York, 2011.

[36] G. Adomian, Solving Frontier Problems of Physics: The Decomposition Method, Kluwer Academic, Dordrecht,

1994.

[37] G. Adomian, On the solution of algebraic equations by the decomposition method. J. Math. Anal. and Appl. 105 (1985) 141-166.

[38] J.-S. Duan, R. Rach, D. Baleanu, A.-M. Wazwaz, A review of the Adomian decomposition method and its applications to fractional differential equations, Commun. Frac. Calc. 3 (2) (2012) 73-99.

[39] G. Lin, General PFG signal attenuation expressions for anisotropic anomalous diffusion by modified-Bloch equations, arXiv:1706.06552v2.

[40] G. Lin, D. Aucoin, M. Giotto, A. Canfield, W. Wen, A.A. Jones, Macromolecules 40 (2007) 1521.

[41] R. Gorenflo, F. Mainardi, Non-Markovian random walks, scaling and diffusion limits, in: O.E. Barndorff-Nielsen (Ed.), Mini-Proceedings of the 2nd MaPhySto Conference on Lévy Processes: Theory and Applications, Department of Mathematics, University of Aarhus, Denmark, 21–25 January 2002 (ISSN 1398-5957), pp. 120–128.

[42] R. Gorenflo, A. Vivoli, Fully discrete random walks for space–time fractional diffusion equations, Signal Process. 83 (2003) 2411 – 2420.

[43] S. Jespersen, R. Metzler, H.C. Fogedby, Lévy flights in external force fields: Langevin and fractional Fokker-Planck equations and their solutions, Phys. Rev. E 59 (1999) 2736.

[44] R. Angelico, A. Ceglie, U. Olsson, G. Palazzo, L. Ambrosone, Phys. Rev. E 74 (2006) 031403.

[45] R. Gorenflo, J. Loutchko, Y. Luchko, Computation of the Mittag-Leffler function $E_{\alpha,\beta}(z)$ and its derivative, Fract. Calc. Appl. Anal. 5 (2002) 491–518.

[46] C. Zeng, Y. Chen, Global Pade approximations of the generalized Mittag-Leffler function and its inverse, Fract. Calc. Appl. Anal. 18 (2015) 1492–1506.

[47] G. Lin, Fractional differential and fractional integral modified-Bloch equations for PFG anomalous diffusion and their general solutions, Physica A **2018**, https://doi.org/10.1016/j.physa.2018.01.008.

[48] C. Choquet, M. C. Neel, Feynman-Kac equation for convection-dispersion with mobile and immobile walkers, Proceedings of the CCT '11Marseille, France, 23 – 27 May 2011, Chaos, Complexity and Transport (2012) 180-188.

[49] C. Choquet and M.C. Neel, Discrete Continuous Dyn. Syst. Ser. S, 7 (2) (2014) 207-238.